# Spin-1/2 *Kagomé*-Like Lattice in Volborthite Cu$_3$V$_2$O$_7$(OH)$_2$·2H$_2$O

Zenji HIROI,[*] Masafumi HANAWA, Naoya KOBAYASHI,[1] Minoru NOHARA,[2] Hidenori TAKAGI,[2] Yoshitomo KATO, and Masashi TAKIGAWA

*Institute for Solid State Physics, University of Tokyo, Kashiwanoha, Kashiwa, Chiba 277-8581, Japan*
[1]*Toda Kogyo Corp., Meijishinkai, Otake, Hiroshima 739-0652, Japan*
[2]*Graduate School of Frontier Sciences, University of Tokyo, Hongo, Bukyo-ku, Tokyo 113-8656, Japan*

**ABSTRACT**

A novel cuprate Volborthite, Cu$_3$V$_2$O$_7$(OH)$_2$·2H$_2$O, containing an $S$-1/2 (Cu$^{2+}$ spin) *kagomé*-like lattice is studied by magnetic susceptibility, specific heat, and $^{51}$V NMR measurements. Signs for neither long-range order nor spin-gapped singlet ground states are detected down to 1.8 K, in spite of large antiferromagnetic couplings of ~ 100 K between Cu spins forming a two-dimensional *kagomé*-like network. It is suggested that Volborthite represents a system close to a quantum critical point between classical long-range ordered and quantum disordered phases.

*E-mail: hiroi@issp.u-tokyo.ac.jp

## §1. Introduction

Geometrical frustration in quantum antiferromagnets (AFMs) tends to stabilize unusual ground states such as a spin glass and a spin liquid instead of classical Néel order. It occurs on various triangle-based lattices like one-dimensional (1D) trestle lattice, two-dimensional (2D) triangular and *kagomé* lattices, and three-dimensional B-site spinel and pyrochlore lattices.[1] In order to reduce total magnetic energy for antiferromagnetically interacting Heisenberg spins on triangles, the compromise arrangement, the so-called 120° state, is realized for the 2D triangular lattice.[2] In contrast, such a compromise arrangement is not stabilized for the more frustrating *kagomé* lattice, because there still remains a degeneracy in propagating the 120° state on a triangle plaquette to neighboring triangles due to corner-sharing.[1] This local degeneracy results in a finite entropy for the classical ground state, and should be lifted by quantum fluctuations. Most theoretical studies have focused on $S$-1/2 Heisenberg antiferromagnets on the *kagomé* lattice, and it has been believed that the ground state is a spin liquid with a finite excitation energy gap $\Delta$.[3-7] However, the physical picture of the ground state as well as the nature of low-lying excitations are still questions under debate. For example, Elstner and Young[5] suggested a spin liquid consisting of short-range singlet dimer pairs with $\Delta \sim 0.25\,J$, where $J$ is the magnitude of pairwise antiferromagnetic (AF) couplings, while Waldtmann et al.[7] claimed a much smaller gap of $0.025\,J < \Delta < 0.1\,J$, implying dominant longer-range correlations. They also insisted that the singlet-triplet gap is filled with nonmagnetic excitations, the origin of which is possibly related to the ground state degeneracy of the classical model.

To clarify the essential feature of the *kagomé* AFMs, we need a real-life material on which a quasi-2D *kagomé* lattice is realized. Unfortunately, however, we have not yet been given such an ideal *kagomé* compound suitable for detailed experimental characterizations. So far well studied are a garnet compound SrCr$_{9-x}$Ga$_{3+x}$O$_{19}$ with Cr$^{3+}$ ($S = 3/2$)[8,9] and the Jarosite family of minerals K$M_3$(OH)$_6$(SO$_4$)$_2$ with $M$ = Cr$^{3+}$ or Fe$^{3+}$.[10-13] In both of them Heisenberg spins form a *kagomé* lattice with strong AF interactions: the Curie-Weiss constant $\Theta$ is -500 K for the former, and -67.5 K (Cr$^{3+}$) or -600 K (Fe$^{3+}$) for the latter. In the case of SrCr$_{9-x}$Ga$_{3+x}$O$_{19}$ no evidence for long-range order (LRO) has been obtained. However, it exhibits a spin-glass transition at 3.5 K instead of a spin gap theoretically expected for the $S = 1/2$ *kagomé* antiferromagnet. It should be noted that this compound suffers crystallographic disorder.[8] On the other hand, Fe Jarosite exhibits LRO below $T_N = 50$ K with the 120° spin structure,[13] while observed in Cr jarosite are a spin-glass transition at $T = 2$ K and a strong dynamic spin fluctuation even at $T = 25$ mK:[11] Tendency toward quantum disorder is strong for spins with smaller spin quantum number. In order to study the exotic ground state due to the geometrical frustration we really long for an $S = 1/2$ *kagomé* compound.

Copper oxides (II) apparently present the most suitable base for $S = 1/2$ quantum AFMs. However, most of them form square-based network, not triangle-based, because they are built up generally with CuO$_4$ square plaquettes connected by their corners or edges. An exceptional triangle-based lattice is found in a natural copper vanadate mineral $Cu_3V_2O_7(OH)_2 \cdot 2H_2O$ called Volborthite. This compound has been known since 18th century,[14] and recently the crystal structure was determined with a synthetically prepared polycrystalline sample by means of Rietveld analyses of powder X-ray and neutron diffraction patterns.[15] It has basically a 2D structure made up from Cu-O(OH) layers pillared by pyrovanadate groups $V_2O_7$ with water molecules embedded between the layers, as perspectively viewed in Fig. 1. Vanadium ions must be in the pentavalent state without $3d$ electrons, judging from the preparation conditions. As described later, the present $^{51}$V NMR experiments confirmed this point. Then, all the copper ions are in the divalent state from charge neutrality, carrying one unpaired electron with $S = 1/2$ per ion. Therefore, the magnetism of Volborthite should substantially come from the Cu-O(OH) layers. In addition, magnetic couplings between adjacent Cu-O(OH) layers may be small because of large separation of ~ 7 Å and relatively negligible superexchange couplings through $V_2O_7$ groups. The Cu sublattice in the layer forms a triangular arrangement, where triangles are connected with each other by their corners as in the *kagomé* lattice. Unfortunately, however, the crystal symmetry is monoclinic, which lacks a three-fold rotation axis, and thus these triangles are not equilateral but slightly elongated to be isosceles triangles. Consequently, one finds a unique, modified $S = 1/2$ *kagomé* lattice in Volborthite. The magnetic properties must be very interesting but have not yet been reported so far. Here we have prepared high quality polycrystalline samples and studied the magnetic properties by means of uniform susceptibility, specific heat, and $^{51}$V NMR.

## §2. Experimental

Polycrystalline samples were prepared by chemical reactions between CuO and $V_2O_5$ in NaOH solutions. First 0.3 mol of CuO powders were dissolved in a $H_2SO_4$ solution at 40°C, and then 0.1 mol of $V_2O_5$ powders were added. After complete disolution the pH of the solution was controlled to be 5.4 by adding a NaOH solution. Finally the solution was heated to 95°C and stirred at this temperature for 2 days. A light green precipitate was obtained which was filtered and dried in air. It was composed of tiny plate-like crystals with mean dimensions ~0.2 μm φ × 0.075 μm. The product was found to be single phase of Volborthite by powder X-ray diffraction. The lattice parameters determined are $a$ = 10.636 Å, $b$ = 5.883 Å, $c$ = 7.224 Å, and $\beta$ = 95.05°, which are in good agreement with those reported previously.[15] Preliminary experiments to grow a "visible" single crystal have been unsuccessful.

Magnetic susceptibility was measured between 1.8 K and 400 K in external fields up to 7 T in a Quantum Design SQUID magnetometer (MPMS-XL). After examinations of several samples prepared in different conditions for pH, temperature, and time, it was found that the Curie tails observed at low temperature in the magnetic susceptibility, which may signal the sample quality, significantly depend on the preparation conditions. The above mentioned condition was selected so as to minimize the Curie tails.

Specific heat was measured by an adiabatic heat-pulse method in a temperature range between 1.8 and 70 K. A sample weighing about 200 mg was attached to a thermometer/heater copper platform with a small amount of Apiezon N grease. Heat capacity of the sample was obtained by subtracting the addenda heat capacity, which had been determined in a separate run without the sample. The resolution of the measurement was about 0.5 % and the absolute accuracy determined from the measurement of a Cu standard was better than 1 %.

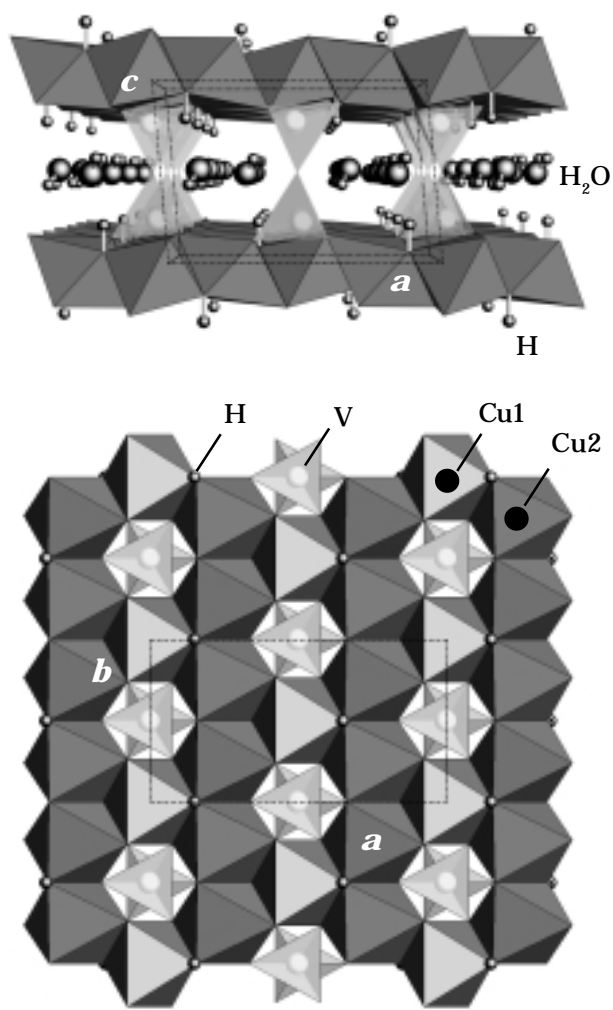

Fig. 1. Crystal structure of Volborthite $Cu_3V_2O_7(OH)_2 \cdot 2H_2O$ drawn on the basis of the structural data reported previously;[15] a perspective, cross-sectional view along the $b$ axis (upper) and a plan-view showing a *kagomé*-like lattice made of Cu ions with $S$ = 1/2. Oxygen ions are not shown but located at the vertices of the polyhedra with Cu or V ions at the center.

NMR spectra of $^{51}$V nuclei were obtained by recording the integrated intensity of the spin-echo signal as a function of magnetic field at a fixed resonance frequency. The spin-lattice relaxation rate ($1/T_1$) was measured by the saturation recovery method. It was not possible to observe the NMR signal at the Cu nuclei presumably because the spin echo decay rate was too large.

## §3. Results

*3.1 Magnetic susceptibility*

The magnetic susceptibility $\chi$ shows a broad, rounded maximum around 20 K without any anomalies indicative of LRO down to 1.8 K, as shown in Fig. 2. The broad maximum implies developing short-range antiferromagnetic correlations below this temperature which is characteristic of low-dimensional AFMs. Such a feature has not been observed in other *kagomé* compounds so far studied. A Curie-like upturn is seen below 5 K in the figure. This may come from "impurity" $Cu^{2+}$ spins possibly created at lattice defects, with the amount estimated to be about 0.5 % by fitting the data measured at 0.1 T between 1.8 and 4 K to the Curie-Weiss law. Ignoring negligibly small contributions from core and Van Vleck paramagnetisms, a large residual spin susceptibility at $T = 0$ of $2.7 \times 10^{-3}$ cm$^3$/mol Cu is estimated. This strongly suggests that the ground state of Volborthite is gapless. No marked hysteretic behaviors are seen depending on whether the sample was zero-field cooled (ZFC) and field cooled (FC) even in small applied fields, as typically shown for 0.01 $T$ in the inset to Fig. 2a. This result excludes the possibility of a spin-glass transition above 1.8 K.

Figure 2b shows an inverse susceptibility versus $T$ plot over a wide temperature range, which exhibits a linear behavior above 200 K. From the slope and its extrapolation to $1/\chi = 0$ the Curie constant $C$ and Weiss temperature $\Theta$ were determined; $C = 0.4713$ cm$^3$/K mol Cu and $\Theta = -115$ K. The effective paramagnetic moment $p_{\mathrm{eff}}$ for Cu was calculated to be 1.95 $\mu_B$, assuming that vanadium ions are nonmagnetic in the pentavelent state. This gives a g-factor of 2.26 which is a typical value for Cu oxides.[16] In the framework of the mean-field theory which considers only $z$ nearest-neighbor ions coupled with superexchange interaction $J$, $\Theta$ is given as $(-zJS(S+1))/(3k_B)$ (The Hamiltonian of the Heisenberg model here is $J\Sigma_{<i,j>}\boldsymbol{S}_i\cdot\boldsymbol{S}_j$). Thus, $k_B\Theta = -J$ for the $S = 1/2$ *kagomé* lattice with $z = 4$, if we ignore the possible geometrical anisotropy in $J$. The obtained $J/k_B$ value of 115 K is similar to those found in other cupric oxides like $CuGeO_3$ containing ~100° Cu-O-Cu bonds. The empirical measure of frustration proposed by Ramirez,[1] $f = -\Theta/T_c$, is larger than 60 in the present compound. For typical *kagomé* compounds $f$ is 150 ($SrCr_8Ga_4O_{19}$) or 39 ($KCr_3(OH)_6(SO_4)_2$).

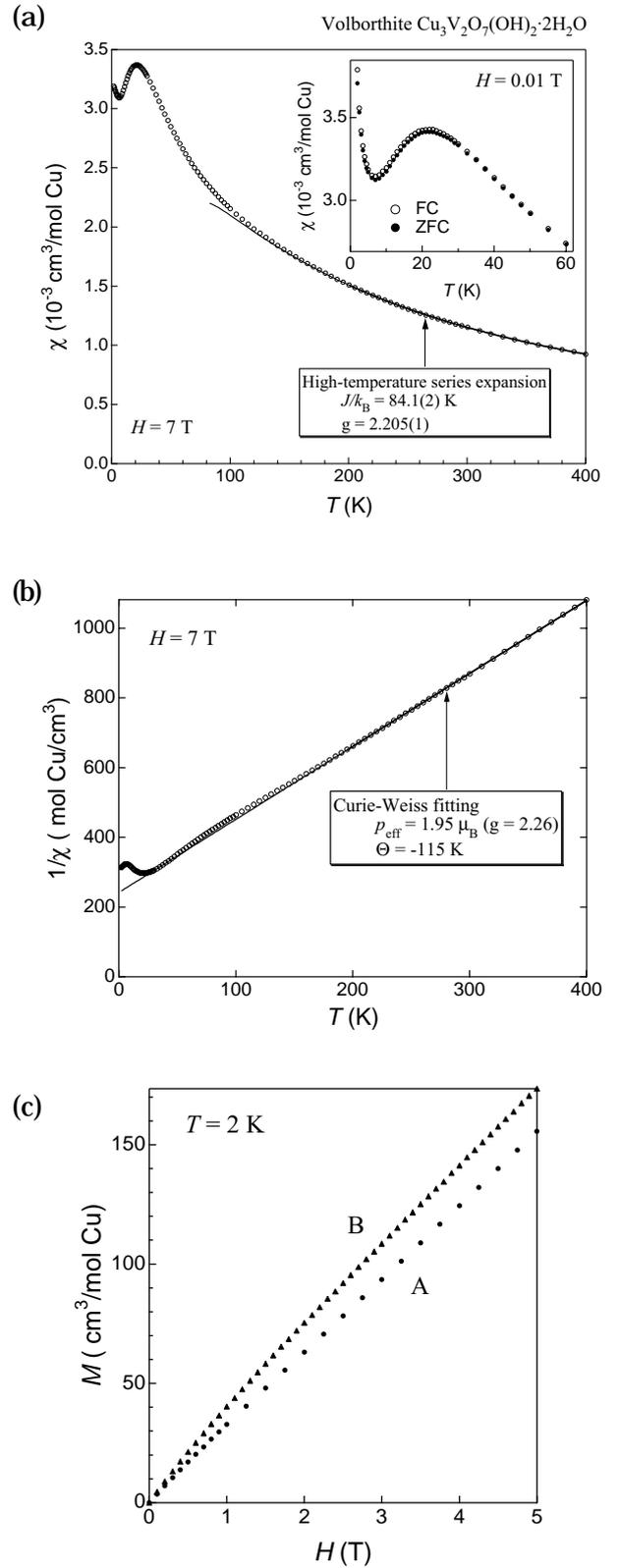

Fig. 2. Magnetic susceptibility $\chi$ (a), inverse susceptibility (b) measured in an applied field of 7 T, and magnetization $M$ at $T = 2$ K plotted against applied field (c). The thick lines on the data points between 200 K and 400 K are fits to the calculation by high-temperature series expansion[5] for the $S = 1/2$ *kagomé*- lattice in (a) and to the Curie-Weiss law in (b). The inset to (a) shows the absence of thermal hysteresis measured in a low field of 0.01 T. The *M-H* curves for two samples A and B are plotted in (c).

At the moment neither analytical forms nor calculated data for the $\chi$-$T$ curve of the *kagomé* antiferromagnet in a wide temperature range is available. We could not fit the data of Fig. 2a assuming 1D Heisenberg chains. If the system is assumed to be quasi-1D, $J$ would be ~ 30 K from the peak-top temperature, which is apparently inconsistent with $J/k_B = -2\Theta$ ~ 230 K ($z = 2$) from the mean-field theory. In order to estimate the $J$ value more accurately within the uniform *kagomé* lattice model, we fitted the $\chi$-$T$ data above 200 K using the result calculated by high-temperature series expansions,[5] as shown in Fig. 2a. Thus estimated parameters are $J/k_B = 84.1(2)$ K and g = 2.205(1). Elstner and Young predicted a low-temperature behavior for $\chi$ from the exact diagonalization of small clusters.[5] Since there is an energy gap in the spin excitation spectrum, the $\chi$ has a maximum at $T \sim J/(6k_B)$ and vanishes exponentially as $T$ approaches zero. Their $\chi$ value and temperature at the maximum are $\chi = 3.37 \times 10^{-3}$ cm$^3$/mol Cu and $T = 14$ K in the case of $J/k_B = 84$ K, which are close to the observed values of $\chi = 3.37 \times 10^{-3}$ cm$^3$/mol Cu and $T = 21$ K, respectively. Therefore, the uniform susceptibility of Volborthite can be described quantitatively with that of the $S = 1/2$ *kagomé* lattice at high temperature, though the low-temperature behaviors below the maximum contrast between the experiments and calculations. It is noted here that the cluster size used in the calculation was not large enough, giving rise to some ambiguity in reproducing low-energy excitations which become dominant at low temperature.

Figure 2c shows the field dependence of magnetization measured at $T = 2$ K on two samples prepared in different conditions. Sample A exhibits the smallest Curie-like upturn in the temperature dependence as shown in Fig. 2a, while sample B has a much larger Curie component. The magnetization curves are convex at low field below ~ 1 T (~2 T) for sample A (B) and become linear at high field. This nonlinear component is obviously due to "impurity" moment which is added to the intrinsic linear term. However, the unusually high saturation fields indicate that the "impurity" spins are not simple free spins but couple significantly with majority spins. As discussed in the NMR section, the entity must be expanded domains with local staggered magnetization created near some kind of defects.

*3.2 Specific heat*

Specific heat $C$ measurements are generally more sensitive to a phase transition than magnetic susceptibility especially in the case of quantum AFMs. The specific heat of Volborthite exhibited no anomalies down to 1.8 K, which evidences absence of LRO above this temperature (Fig. 3). The anomaly seen at 9 K is an experimental artifact. It was not easy to extract a magnetic contribution $C_m$ from the measured data, because a nonmagnetic isomorph is not available at present from which the lattice contribution could be estimated. We fitted the data at 50 - 70 K to the simple Debye model assuming negligible magnetic contributions at this high-temperature range. The estimated Debye temperature was 320 K. Thus determined lattice contribution $C_D/T$ is plotted with the broken line in Fig. 3. A magnetic contribution is determined as $C_m = C - C_D$ and is also plotted with solid circles in the figure. Integrating $C_m/T$ between 1.8 K and 60 K, we find a value of 4.1 J/mol K which is about 30 % smaller than the total magnetic entropy (Rln2) for $S = 1/2$. The discrepancy is most likely due to crudeness of the estimation of the lattice contribution particularly at high temperature. To be noted here is that $C_m/T$ seems to show a broad maximum at 20-25 K and then rather steep decrease below 3 K. Alternatively, one can say that there is a second peak or shoulder around 3 K. The first maximum must be ascribed to short-range AF ordering, because its temperature coincides

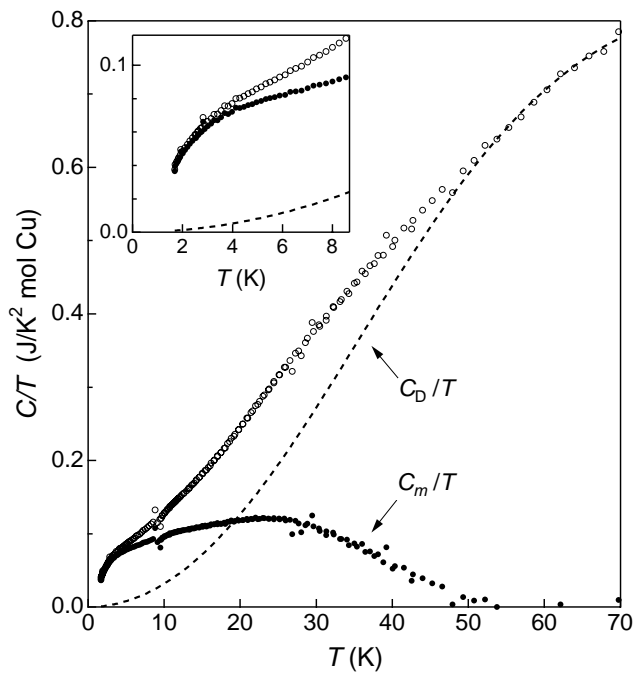

Fig. 3. Specific heat of Volborthite. The open circles show the raw data, while the magnetic contribution estimated is shown with closed circles. The estimated, Debye-type lattice contribution is shown with the broken line. The inset is an enlargement at low temperature.

with that of the broad maximum observed in the susceptibility.

Elstner and Young predicted theoretically that the $C/T$ of the $S = 1/2$ *kagomé* AFM exhibits a broad maximum at $J/(3k_B)$ followed by a second sharp peak at $T/k_B \sim J/10$ ( In their original paper $C$ was shown to have a broad maximum at $T/k_B = (2/3)J$, while we translated it into that $C/T$ does at $T/k_B = (1/3)J$.).[5] They discussed that quantum tunneling removes the large degeneracy of the classical ground state, leading to many low-lying (but split) states for the quantum case. This would give additional structures to the specific heat at low temperature. The second peak is presumably associated with this quasi-degenerate low-lying states. Recently, such a second peak was detected experimentally at very low temperature in the second layer of $^3$He physisorbed on graphite which was believed to realize another class of *kagomé* lattice.[17] In the case of the present compound one would expect a broad maximum at 28 K and the second peak at 8.4 K, assuming $J/k_B = 84$ K. Therefore, the overall feature of the specific heat for Volborthite is not far away from expected for the *kagomé* model. However, the second peak observed is much smaller than theoretically expected. It is very important to clarify how the system approaches the zero-entropy state as $T \to 0$. An unusual $T^2$ dependence in $C$ was reported for $SrCr_{9-x}Ga_{3+x}O_{19}$.[18] Further specific heat measurements at lower temperature are planned.

*3.3 NMR*

The $^{51}$V powder NMR spectra obtained at different temperatures and frequencies (magnetic fields) are shown in Fig. 4. The spectrum at 89 MHz and $T = 230$ K (Fig. 4(a)) shows asymmetric line shape with several distinct peaks resulting from combined effects of the anisotropic magnetic hyperfine interaction and the electric quadrupole interaction from spin 7/2 nuclei. The spectrum becomes more symmetric at lower frequencies (Fig. 4(b)), where the quadrupole interaction is dominant. Since it is apparent from the spectrum in 4(a) that the magnetic hyperfine shift ($K$) is dominantly isotropic, $K$ is determined from the main peak of the spectrum at 89 MHz and plotted against temperature in Fig. 5. The temperature dependence of $K$ is similar to the susceptibility. When $K$ is plotted against $\chi$, it reveals approximately a linear relation above 40 K, $K = A\chi/N_A\mu_B + K_0$, yielding the values of the hyperfine coupling constant $A = 0.66$ T/$\mu_B$. Since the core polarization hyperfine field from the on-site 3$d$ electrons is of the order of -10 T/$\mu_B$, the small value of $A$ indicates that vanadium ions are non-magnetic (pentavalent) and the hyperfine field at the V nuclei originates from neighboring Cu spins.

The NMR spectrum becomes significantly broadened at lower temperatures as shown in Fig. 4(c). The spectrum consists of the relatively narrow central peak and the broad background. The full widths of the spectra at 1/2, 1/10, and 1/20 of the peak height obtained at 12.8 MHz (near 1.1 T) are plotted against temperature in the main panel of Fig. 6. Although the 1/2 width, which characterizes the central peak, increases only modestly, both the 1/10 and 1/20 widths characterizing the background shows steep increase below 10 K. The field dependence of these widths at $T = 1.7$ K is shown in the inset. While tendency for saturation can be seen at high fields, all widths appear to vanish at zero field, indicating that the line broadening is due to paramagnetic (field induced) moments. The pronounced anomaly in the line width is contrasting to the only modest increase of the bulk susceptibility below 10 K. This indicates that the symmetric background is due to local staggered (oscillating) magnetization near impurities while nuclei far from impurities contribute to the central peak. Indeed this shape is quite similar to that observed in $Sr_2CuO_3$, a spin 1/2 1D Heisenberg system, in which local staggered magnetization is induced by magnetic field near chain ends.[19] Similar spectrum due to impurity-induced local staggered magnetization has been observed in a number of 1D antiferromagnets with spin gap, where unpaired free spins are created near impurities.[20, 21] We therefore suppose that also in our two dimensional material local staggered magnetization is created near some kinds of defects.

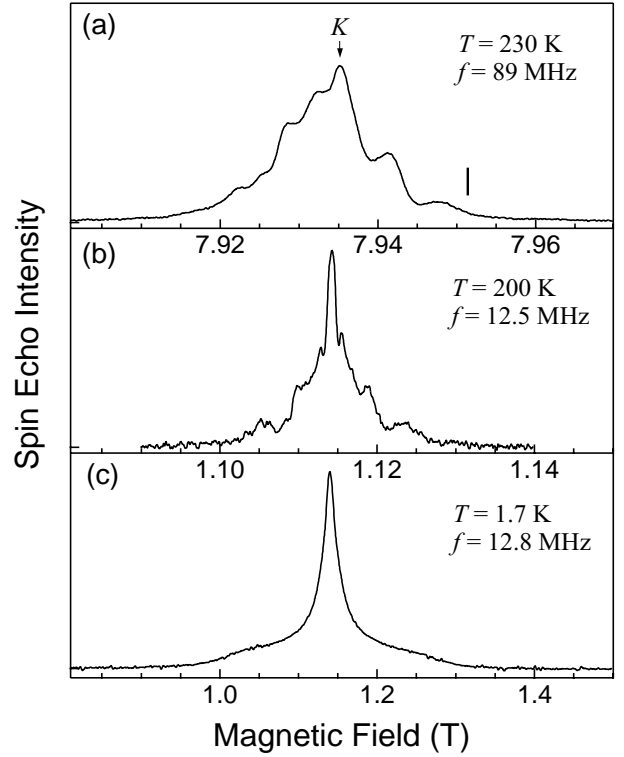

Fig. 4. $^{51}$V powder NMR spectra at different temperatures and frequencies. The vertical bar in (a) indicates the resonance field in diamagnetic substances. The magnetic hyperfine shift is determined from the peak field at $f = 89$ Hz marked by arrow.

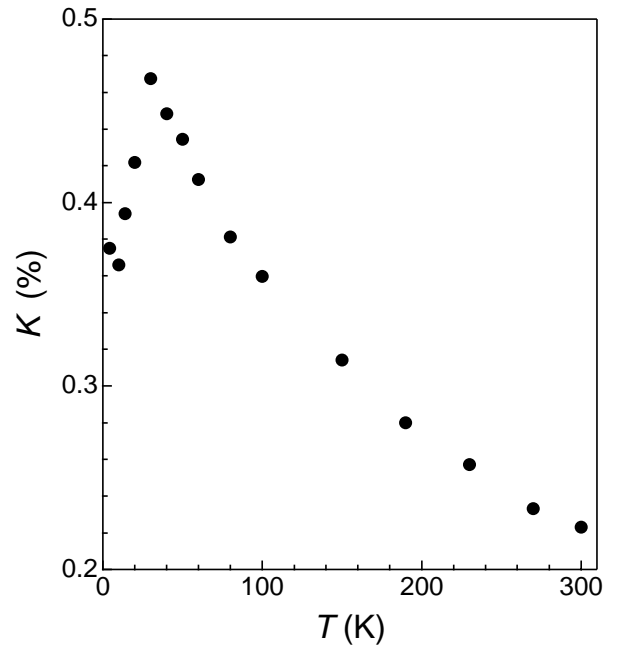

Fig. 5. Magnetic hyperfine shift $K$ determined from the peak field at 89 MHz plotted against temperature.

The spin-lattice relaxation rate ($1/T_1$) measured at 7.9 T is plotted against temperature in Fig. 7. The recovery of the nuclear magnetization after the inversion pulse is exponential over two decades above 60 K. However, deviation from exponential functions becomes substantial at lower temperatures indicating large inhomogeneous distribution of $1/T_1$.

Generally, $1/T_1$ in magnetic insulators is temperature independent in the high temperature limit ($T \gg J/k_B$). If the system has long range Néel order, $1/T_1$ diverges towards $T_N$ due to the critical slowing down of spin fluctuations. On the other hands, if the system has singlet ground state with a spin gap, $1/T_1$ falls exponentially at low temperatures. Thus the observed temperature dependence of $1/T_1$ is quite peculiar since 1) it decreases gradually with decreasing temperature even when the temperature is much larger than the exchange $J$ and 2) it appears to approach a finite value as $T$ goes to zero, although the intrinsic behavior is masked at low temperatures due to inhomogeneous distribution in $1/T_1$. The second feature is expected when the system is near the quantum critical point in two dimension between Néel ordered and disordered ground states.[22]

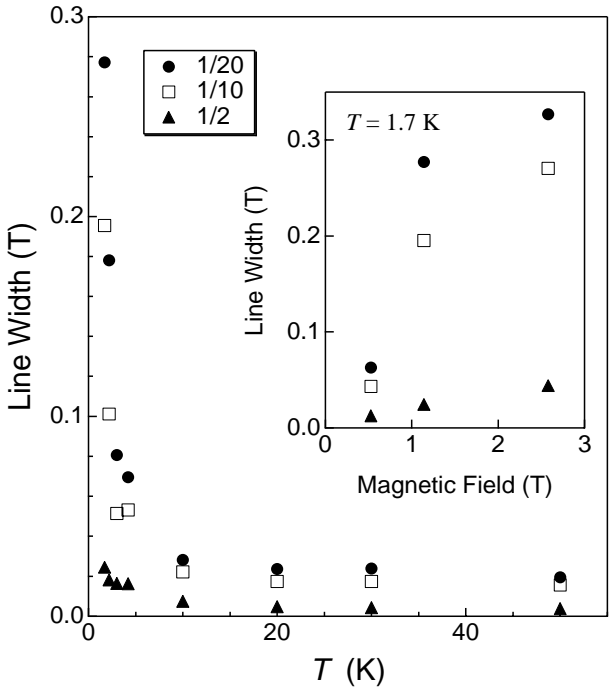

Fig. 6. Temperature dependence of the full width at 1/2, 1/10, and 1/20 of the peak height obtained at 12.8 MHz (near 1.1 T). The inset shows the field dependence of these widths at 1.7 K.

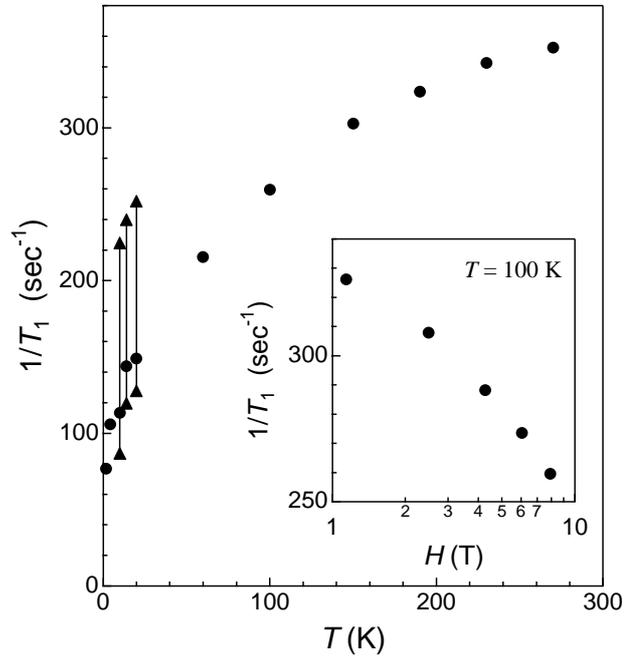

Fig. 7. Temperature dependence of $1/T_1$ measured at 89 MHz. The recovery curve of the nuclear magnetization can be fit to a single exponential function above 60 K. Between 10 and 20 K, it was fit to a sum of two exponential functions with different relaxation rates as shown by triangles. Below 10 K, the distribution of $1/T_1$ is almost continuous and only the smallest component of the relaxation rates are shown. In the inset is shown the magnetic field dependence of $1/T_1$ at $T = 100$ K.

Assuming that a V nucleus is coupled to six nearest neighbor Cu spins by isotropic hyperfine interaction, we can estimate the value of $1/T_1$ in the high temperature limits where there is no spatial spin correlation

$$\frac{1}{T_{1\infty}} = \frac{\sqrt{2\pi}\gamma_N^2 g^2 A^2 S(S+1)}{3z_1\omega_e}, \qquad (3.1)$$

where $\gamma_N = 2\pi\times1.1193\times10^7$ sec$^{-1}$T$^{-1}$ is the nuclear gyromagnetic ratio, $z_1 = 6$ and $\omega_e$ is the exchange frequency given as

$$\omega_e = \frac{J}{\hbar}\sqrt{\frac{2zS(S+1)}{3}} \qquad (3.2)$$

with $z = 4$. If we use the values $J/k_B = 84$ K and $A = 0.66$ T/$\mu_B$ estimated above, we obtain $1/T_{1\infty} = 58$ sec$^{-1}$, which is significantly smaller than the experimental value at room temperature. This implies that spin fluctuations are slower than what is expected for ordinary Heisenberg antiferromagnets. We also found that $1/T_1$ depends substantially on magnetic field as shown in the inset to Fig. 7. It is likely that such field dependence is due to spin diffusion in two dimension. The slow hydrodynamic spin fluctuations at long wavelength in 2D leads to ln$H$ singularity. Although the measured $1/T_1$ deviates from ln$H$ behavior at low fields, this could be due to three dimensional coupling which cuts off the 2D spin diffusion.

§4. Discussion

In order to get more insight on the superexchange couplings and their anisotropy on the kagomé-like net of Volborthite, let's look at in more detail the crystal structure. Based on the structural data given by Lafontaine et al.[15] the local atomic coordination of O and OH ions around one triangle of Cu ions is depicted in Fig. 8. Every Cu ion is surrounded octahedrally by two OH ions pointing in opposite directions and four planar O ions. However, there are two crystallographically distinguished Cu sites; Cu1 in a compressed octahedron and Cu2 in an elongated one. A Cu1 octahedron is connected with four Cu2 octahedra by sharing their edges, while a Cu2 decahedron with two Cu1 and two Cu2 octahedra. Since there is a mirror plane at the Cu1 site perpendicular to the b axis, two of three superexchange pathways on the Cu triangle should be identical, denoted as $J_1$ between Cu1 and Cu2. On the other hand, the third path between two neighboring Cu2 sites, denoted as $J_2$, can be different from the other. Note that both the superexchange couplings may occur through nearly 90° Cu-O-Cu and Cu-OH-Cu bonds which are generally sensitive to the bond angle. The actual angles in Volborthite are 83.7° and 105.6° for $J_1$ (Cu1-Cu2), and 91.5° and 101.1° for $J_2$ (Cu2-Cu2), respectively. Moreover, the arrangement of the $dx^2-y^2$ orbital must be taken into account. Although it is practically difficult to evaluate the magnitude of superexchange couplings, these significant differences in the local structure suggest considerable anisotropy in $J$s. The value of 84 K estimated from fitting the susceptibility data might correspond to an average of $J_1$ and $J_2$ ($3J = 2J_1 + J_2$). The Cu sublattice in Volborthite is considered to be an $S = 1/2$ kagomé-like network composed of isosceles triangles, as illustrated in Fig. 8. This structure still retains frustration, though it is reduced compared with the ideal kagomé lattice.

All the experimental facts given in the present study have indicated that Volborthite exhibits neither long-range order nor spin-glass order down to 1.8 K in spite of rather large antiferromagnetic couplings ($J/k_B = 84$ K and $\Theta = -104$ K). On the other hand, it is also revealed that there are no clear sign for opening a gap above 1.8 K that signifies a spin liquid ground state. However, either of them should appear in the ordinary spin system as $T$ approaches zero, in order to reach the zero-entropy state. We consider three possible scenarios to understand our experimental data.

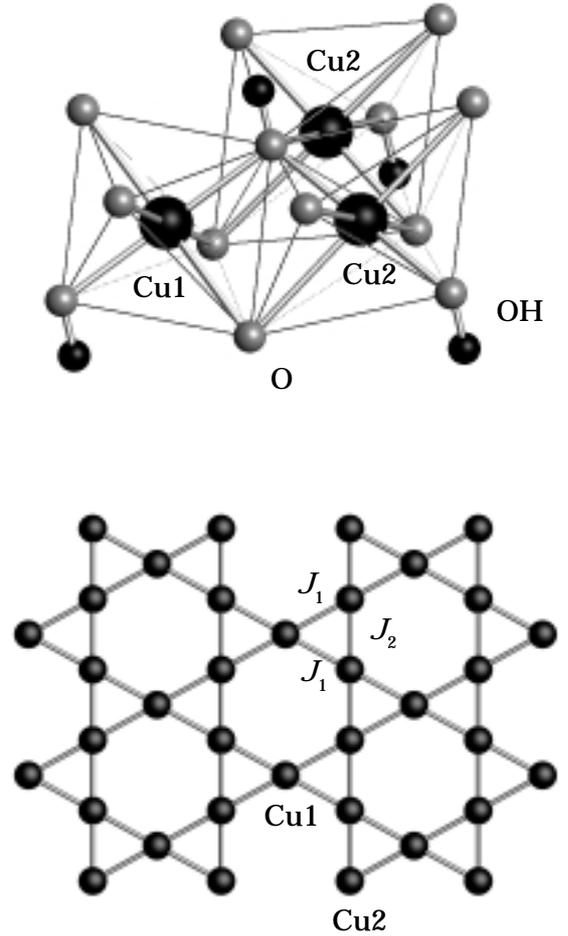

Fig. 8. Local arrangement of Cu-O$_4$(OH)$_2$ octahedra around one Cu1-Cu2-Cu2 triangle (upper) and the schematic representation of the modified kagomé network made of isosceles triangles seen in Volborthite (lower). In the lower figure balls represent Cu ions with $S = 1/2$ with the others omitted for clarity.

The first one assumes that the critical temperature for LRO is lower than 1.8 K. However, since no critical behaviors toward LRO were detected in $1/T_1$ and $C$ above 1.8 K, the critical temperature would be much lower. A recent μSR study on our sample indicated that Cu spins are not statically frozen but dynamically fluctuating even at $T$ = 20 mK.[23] Therefore, the possibility of a classically ordered ground state would be unrealistic.

The second scenario assumes that there is certainly a small gap, $\Delta/k_B \ll 1.8$ K. Previous theoretical studies seem to converge on the point that an ideal $S = 1/2$ *kagomé* AFM has an energy gap of $0.025J \sim 0.25J$, which corresponds to $\Delta/k_B = 2.1$ K $\sim$ 21 K for $J/k_B = 84$ K.[5,7] It would be a possible case if the true size of the gap is smaller than predicted or the anisotropy in $J$ reduces the gap. Another important point to be noted is the appearance of Curie-like contributions at low temperature observed in $\chi$ and $K$. Moreover, the linewidth of the NMR spectrum increases rapidly below 10 K. A similar broadening was also detected in the ESR peak measured by Okubo et al.[24] These facts remind of an impurity-induced staggered polarization around defects recently studied well in the spin-gapped systems doped with impurities, $(Cu,Zn)(Ge,Si)O_3$[25] and $Sr(Cu_{1-x}Zn_x)_2O_3$.[26] It is likely that the sample studied here may have already contained a small amount of impurities or defects, and thus a sign of the spin gap is masked in the proximity of impurity-induced AF order. A single crystal of high quality would be necessary for further characterization.

In general, the size of the spin gap is inverse-proportional to the spin-spin correlation length $\xi$: A small gap means a large correlation length. In the limit of $\Delta \to 0$, $\xi$ goes to infinity, where quantum critical phenomena are expected. This unusual case is actually realized in 1D HAF chains where neither LRO nor spin-gapped state is seen as $T \to 0$. In contrast a 2D analogue is not known. The present material with modified *kagomé* lattice may be located very close to such a quantum critical point. This third scenario can give reasonable explanations for the present experimental results. Especially, the fact that $1/T_1$ approaches a finite value as $T$ goes to zero supports this conclusion.

## §4. Conclusions

We have investigated the magnetic properties of Volborthite $Cu_3V_2O_7(OH)_2 \cdot 2H_2O$ which is considered as a candidate for an $S = 1/2$ *kagomé* antiferromagnet by means of magnetic susceptibility, specific heat, and $^{51}V$ NMR measurements. The results have revealed that the compound exhibits neither long-range order nor spin-gapped singlet ground states down to 1.8 K. It is suggested that the system lies in the quantum critical regime of two dimension.

## Acknowledgments

This research was supported by a Grant-in-Aid for Scientific Research on Priority Areas (A) given by The Ministry of Education, Culture, Sports, Science and Technology, Japan, and by the CREST program (Core Research for Evolutional Science and Technology) of the Japan Science and Technology Corporation. We wish to thank Dr. H. Ohta for his enlightening discussions.

## References

1) A. P. Ramirez: Annu. Rev. Mater. Sci. **24** (1994) 453.
2) D. A. Huse and V. Elser: Phys. Rev. Lett. **60** (1988) 2531.
3) C. Zeng and V. Elser: Phys. Rev. B **42** (1990) 8436.
4) R. R. P. Singh and D. A. Huse: Phys. Rev. Lett. **68** (1992) 1766.
5) N. Elstner and A. P. Young: Phys. Rev. B **50** (1994) 6871.
6) T. Nakamura and S. Miyashita: Phys. Rev. B **52** (1995) 9174.
7) C. Waldtmann, H.-U. Everts, B. Bernu, C. Lhuillier, P. Sindzingre, P. Lecheminant and L. Pierre: Eur. Phys. J. B **2** (1998) 501.
8) X. Obradors, A. Labarta, A. Isalgue, J. Tejada, J. Rodriguez and M. Pernet: Solid State Commu. **65** (1988) 189.
9) A. P. Ramirez, G. P. Espinosa and A. S. Cooper: Phys. Rev. B **45** (1992) 2505.
10) M. Takano, T. Shinjo and T. Takada: J. Phys. Soc. Jpn **13** (1971) 1.
11) A. Keren, K. Kojima, L. P. Le, G. M. Luke, W. D. Wu, Y. J. Uemura, M. Takano, H. Dabkowska and M. J. P. Gingras: Phys. Rev. B **53** (1996) 6451.
12) S.-H. Lee, C. Broholm, M. F. Collins, L. Heller, A. P. Ramirez, C. Kloc, E. Bucher, R. W. Erwin and N. Lacevic: Phys. Rev. B **56** (1997) 8091.
13) T. Inami, M. Nishiyama, S. Maegawa and Y. Oka: Phys. Rev. B **61** (2000) 12181.
14) C. Guillemin: Bull. Soc. fran. Miner. Crist. **LXXIX** (1956) 219.
15) M. A. Lafontaine, A. L. Bail and G. Férey: J. Solid State Chem. **85** (1990) 220.
16) D. C. Johnston, in *Handbook of magnetic materials*, edited by K. H. J. Buschow (Elsevier, 1997), Vol. 10, p. 1.
17) K. Ishida, M. Morishita, K. Yawata and H. Fukuyama: Phys. Rev. Lett. **79** (1997) 3451.
18) A. P. Ramirez, G. P. Espinosa and A. S. Cooper: Phys. Rev. Lett. **64** (1990) 2070.
19) M. Takigawa, N. Motoyama, H. Eisaki and S. Uchida: Phys. Rev. B **55** (1997) 14129.
20) G. B. Martins, M. Laukamp, J. Riera and E. Dagotto: Phys. Rev. Lett. **78** (1997) 3563.
21) F. Tedoldi, R. Santachiara and M. Horvatic: Phys. Rev. Lett. **83** (1999) 412.
22) A. V. Chubkov, S. Sachdev and J. Ye: Phys. Rev. B **49**

(1994) 11919.
23) T. Ito, A. Fukaya and Y. J. Uemura: private communication.
24) S. Okubo, H. Ohta, K. Hazuki, T. Sakurai, N. Kobayashi and Z. Hiroi: Physica B **294-295** (2001) 75.
25) S. B. Oseroff, S.-W. Cheong, B. Aktas, M. F. Hundley, Z. Fisk and J. L.W. Rupp: Phys. Rev. Lett. **74** (1995) 1450.
26) M. Azuma, Y. Fujishiro, M. Takano, T. Ishida, K. Okuda, M. Nohara and H. Takagi: Phys. Rev. B **55** (1997) R8658.